\documentclass{article}
\usepackage{graphicx}
\usepackage{amsmath}
\usepackage{authblk}
\usepackage{color}
\usepackage{natbib}
\usepackage{enumitem}
\usepackage{hyperref}

\begin{document}


\title{Geostatistical models for zero-inflated data and extreme values}


\author[1]{Soraia Pereira \thanks{ sapereira@fc.ul.pt}}

\author[2]{Raquel Menezes}

\affil[1]{Centro de Estatística e Aplicações, Faculdade de Ciências da Universidade de Lisboa, Portugal}

\affil[2]{Universidade do Minho e Centro de Estatística e Aplicações, Faculdade de Ciências da Universidade de Lisboa, Portugal}

\author[3]{Maria Manuel Ang\'{e}lico}

\affil[3]{Instituto Português do Mar e da Atmosfera (IPMA), Lisboa, Portugal.}

\author[4]{Tiago Marques}

\affil[4]{CREEM - University of St Andrews, UK, and Departamento de Biologia Animal, Centro de Estatística e Aplicações, Faculdade de Ciências da Universidade de Lisboa, Portugal}

\date{}

\maketitle


\begin{abstract}
	
\noindent Understanding the spatial distribution of animals, during all their life phases, as well as how the distributions are influenced by environmental covariates, is a fundamental requirement for the effective management of animal populations. Several geostatistical models have been proposed in the literature, however often the data structure presents an excess of zeros and extreme values, which can lead to unreliable estimates when these are ignored in the modelling process.

\noindent To deal with these issues, we propose a point-referenced zero-inflated model to model the probability of presence together with the positive observations and a point-referenced generalised Pareto model for the extremes. Finally, we combine the results of these two models to get the spatial predictions of the variable of interest. We follow a Bayesian approach and the inference is made using the package R-INLA in the software R.

\noindent Our proposed methodology was illustrated through the analysis of the spatial distribution of sardine eggs density (eggs/$m^3$). The results showed that the combined model for zero-inflated and extreme values improved the spatial prediction accuracy.

\noindent Accordingly, our conclusion is that it is relevant to consider the data structure in the modelling process. Also, the hierarchical model considered can be widely applicable in many ecological problems and even in other contexts.

\end{abstract}







\newcommand{\bbeta}{\boldsymbol{beta}}

\newcommand{\bz}{{\bf z}}
\newcommand{\bx}{{\bf x}}
\newcommand{\bs}{{\bf s}}
\newcommand{\bw}{{\bf w}}
\newcommand{\bW}{{\bf W}}
\newcommand{\by}{{\bf y}}
\newcommand{\btheta}{{\boldsymbol\theta}}
\newcommand{\btbeta}{{\boldsymbol\beta}}



\section{Introduction}\label{int}

The ability to predict where, and when, a species will be, and at what densities it will occur when present, is fundamental knowledge to effective management and conservation of wild species. Specific applications might include which areas to choose when doing the translocation of a threatened species \citep{Draper2019}, how to design reserves for species of conservation concern  \citep{Veloz2015}. Species distribution models (SDMs) are typically used to link data on species presence and abundance to spatially (and temporally) indexed covariates, allowing for model based predictions over space and time \citep{Elith2009,Sofaer2019}. 

This is also a geostatistical problem, where the main objective is the prediction of a variable of interest over a domain, based on values observed at a limited number of points. Kriging is a classical approach to spatial prediction in such point-referenced data setting \citep{Diggle2002}. However, inference on such models is not straightforward due to the dense covariance matrices. That problem is known in the literature as big n problem \citep{AdrianCentreforExplorationTargeting2015}. 
To overcome the computational costs, \cite{Lindgren2011} proposed a new approach based on stochastic partial differential equation (SPDE) models. The idea is to approximate the Gaussian field by a Gaussian Markov random field, a discretized version. This approximation can be easily implemented using the integrated nested Laplace approximation (INLA) approach \citep{Rue2009}. 


Some SDMs based on that approach have been proposed in the literature, to solve ecological problems. Typically based on some sort of regression modelling approach, these models must account for the characteristics of the data. Common difficulties, very often present but generally ignored, include too many zeros in the data \citep{Minaya2018},  the need to account for spatial and/or temporal autocorrelation \citep{Dormann2007}, extrapolation in a multivariate covariate space \citep{Yates2018}, and extreme values. 

Thus, here we propose to extend those models to deal with zero-inflated data, and extreme values. An illustration of the proposed methods is done for a spatial analysis of sardine eggs data.


In exploited small pelagic fish populations, recruitment success and sustained healthy stock abundances are determined by fishing pressure regulation, but also by a set of natural, biological and environmental factors. The conditions that influence the survival (mortality) of the initial life stages, egg and larvae, are particularly relevant for the population success as during these phases fish are subjected to very high mortality rates. During the pelagic egg life, which in the sardine (\textit{Sardina pilchardus, Walbaum, 1792}) lasts for around 3 to 5 days, individuals are exposed to predation, infections, water currents and water physical/chemical characteristics (e.g. temperature, salinity). Research on sardine spatial egg distribution and on the environmental components that structure said distribution are important to understand the frequent fluctuations in the abundance of this commercially important species.


In this paper, we start by describing the proposed geostatistical zero-inflated model in section 2.1, and the geostatistical generalised Pareto model to deal with extreme values in section 2.2. The illustration of the methods proposed, applied to a dataset of sardine eggs along the Portuguese coast, is presented in section 2.3, where we discuss the results. Finally, in section 2.4, we present a plug-in method for the joint modelling of the excess zeros and extreme values. A discussion is presented in section 3.

\section{Materials and Methods}\label{mat}

\subsection{Geostatistical zero-inflated model}\label{zero}
\label{ZImodel}

Often ecological data present an excess of zeros
that is very often overlooked when a distribution is chosen to fit the data. 

Let us assume that our data are the locations where the number of a specific species is detected and recorded, and the respective records. Let us suppose that the data have a high percentage of zeros and, consequently, the most common distributions are not adequate.

Here we propose to adopt a geostatistical zero-inflated model which is based on a bivariate model to fit the probability of presence together with the positive observations. This type of model is not new in the fisheries setting, and is sometimes referred to as a delta-gamma model \citep[e.g.][]{Lecomte2013}.

Let us denote $Z(s)$ as the variable which takes the value 0 if there are no presence at location s and 1 otherwise, and  $Y(s)$ the positive observations at location $s$. The hierarchical structure of the proposed model, henceforth denoted by Model I, can be represented as

\begin{enumerate}
	
	\item Data$|$Parameter
	\begin{equation}
	Z(s) \sim \text{Bernoulli}(p(s)) 
	\end{equation}
	\begin{equation}
	Y(s) \mid Z(s)=1 \sim \text{Gamma}(a(s),b(s))
	\end{equation}
	\item Parameter$|$Hyperparameters
	\begin{equation}
	\label{modBIN}
	logit(p(s))=\alpha_1+  \sum_{m=1}^{M_1} \beta_{1,m} X_{1,m}(s) +W_1(s) 
	\end{equation}
	\begin{equation}
	\label{modGAMMA}
	log(a(s)/b(s))=\alpha_2+  \sum_{m=1}^{M_2} \beta_{2,m} X_{2,m}(s) +W_2(s)
	\end{equation}
	for $i=1,2$ and $m=1,\dots,M_i$, $\theta_i=\{ \alpha_i, \beta_{i,m}\}$ are the model parameters, $\{X_{i,m}(s)\}$ are the covariates, and $W_i(s)$ are two independent Gaussian random fields. 
	
	\item Hyperparameters
	\begin{equation}
	\alpha_i \sim N(0,1000), \ i=1,2
	\end{equation}
	\begin{equation}
	\beta_{i,m} \sim N(0,1000), \ i=1,2; \ m=1,...,M_i
	\end{equation}
	
\end{enumerate}

The INLA methodology uses a computational mesh for representing the latent Gaussian field. 

Following \cite{Lindgren2011}, it is assumed the following approximation

\begin{equation}
W(s) \approx \sum_{j=1}^{N} w_j \psi_j(s)
\end{equation}
\noindent where $N$ is the number of the mesh nodes, $w=(w_1,w_2,...,w_N)^T$ is a multivariate random vector, representing a Gaussian Markov random field (GMRF) and $\{\psi_j\}_{j=1}^N$ are the selected base functions defined for each mesh node: $\psi_j$ is 1 at mesh node $j$ and 0 in all other mesh nodes. $w$ is chosen so that the distribution of $W(s)$ approximates the distribution of the solution to the SPDE.

\subsection{Geostatistical model for extremes}\label{extr}


Even more often ignored than the excess of zeros in the ecological literature, are the extreme values. However, failing to model them appropriately could lead to problems. As an example, these extreme values might correspond to a small percentage of the observed values, but if you are interested in estimating a total over a given area, they could make up for a large proportion of said total.

Here we propose a geostatistical generalised Pareto model for the extremes. In particular, we propose a bivariate model to model the probability of exceeding a specified threshold together with the exceedances above that threshold. Let us denote $Z^*(s)$ the variable which takes the value 0 if the observation is lower than the specified threshold at location $s$, and 1 otherwise, and $Y^*(s)$ the exceedances above the threshold.

For the threshold choice, we propose to look at the mean residual life plot, which is obtained by plotting the threshold $u$ against the sample mean excess (mean exceedances - $u$), for a range of $u$. The choice should respect a reasonable balance between precision, which is higher for small thresholds, and bias, which is higher for large thresholds. Following \cite{Coles2001}, the plot should be approximately linear, above the {\em ideal} threshold.

The proposed model for the extremes, henceforth denoted by Model II, can be represented by the following hierarchical structure

\begin{enumerate}
	
	\item Data$|$Parameter
	\begin{equation}
	Z^*(s) \sim \text{Bernoulli}(p^*(s)) 
	\end{equation}
	\begin{equation}
	Y^*(s) \mid Z^*(s)=1 \sim \text{GP}(\sigma,\xi)
	\end{equation}
	\item Parameter$|$Hyperparameters
	\begin{equation}
	\label{modBIN2}
	logit(p^*(s))=\alpha^*_1+  \sum_{m=1}^{M_1} \beta^*_{1,m} X_{1,m}(s) +W^*_1(s) 
	\end{equation}
	\begin{equation}
	\label{modGP}
	log(q_{0.5})=\alpha^*_2+  \sum_{n=1}^{M_2} \beta^*_{2,m} X_{2,m}(s) +W^*_2(s)
	\end{equation}	
	for $i=1,2$ and $m=1,\dots,M_i$, $\theta^*_i=\{ \alpha_i, \beta_{i,m}\}$ are the model parameters, $\{X_{i,m}(s)\}$ are the covariates, $W^*_i(s)$ are two independent Gaussian random fields, and $q_{0.5}$ is the 0.5 quantile. The scale parameter $\sigma$ is a function of $q_{0.5}$ and $\xi$, such that $\sigma= \frac{\xi q_{0.5}}{0.5^{-\xi}-1}$. 
	
	\item Hyperparameters
	\begin{equation}
	\alpha^*_i \sim N(0,1000), \ i=1,2
	\end{equation}
	\begin{equation}
	\beta^*_{i,m} \sim N(0,1000), \ i=1,2; \ m=1,...,M_i
	\end{equation}
	
\end{enumerate}

\subsection{Sardine eggs data example}\label{sard}


The sardine egg data used in these analyses were gathered by Instituto Português do Mar e da Atmosfera (IPMA) in 2018 (28 April - 30 May). IPMA conducts annually a spring acoustics-trawl survey (PELAGO series, PNAB-EU/DCF-FEAMP) with the aim of assessing the stocks of the main small pelagic fish, in particular sardine, in the western and southern area of the Atlantic Iberian Peninsula,  from Cape Trafalgar, in Cadiz Bay, to the northern Portugal-Spain border (Figure 1).

\begin{figure}[h]
	\centering
	\includegraphics[width=7cm]{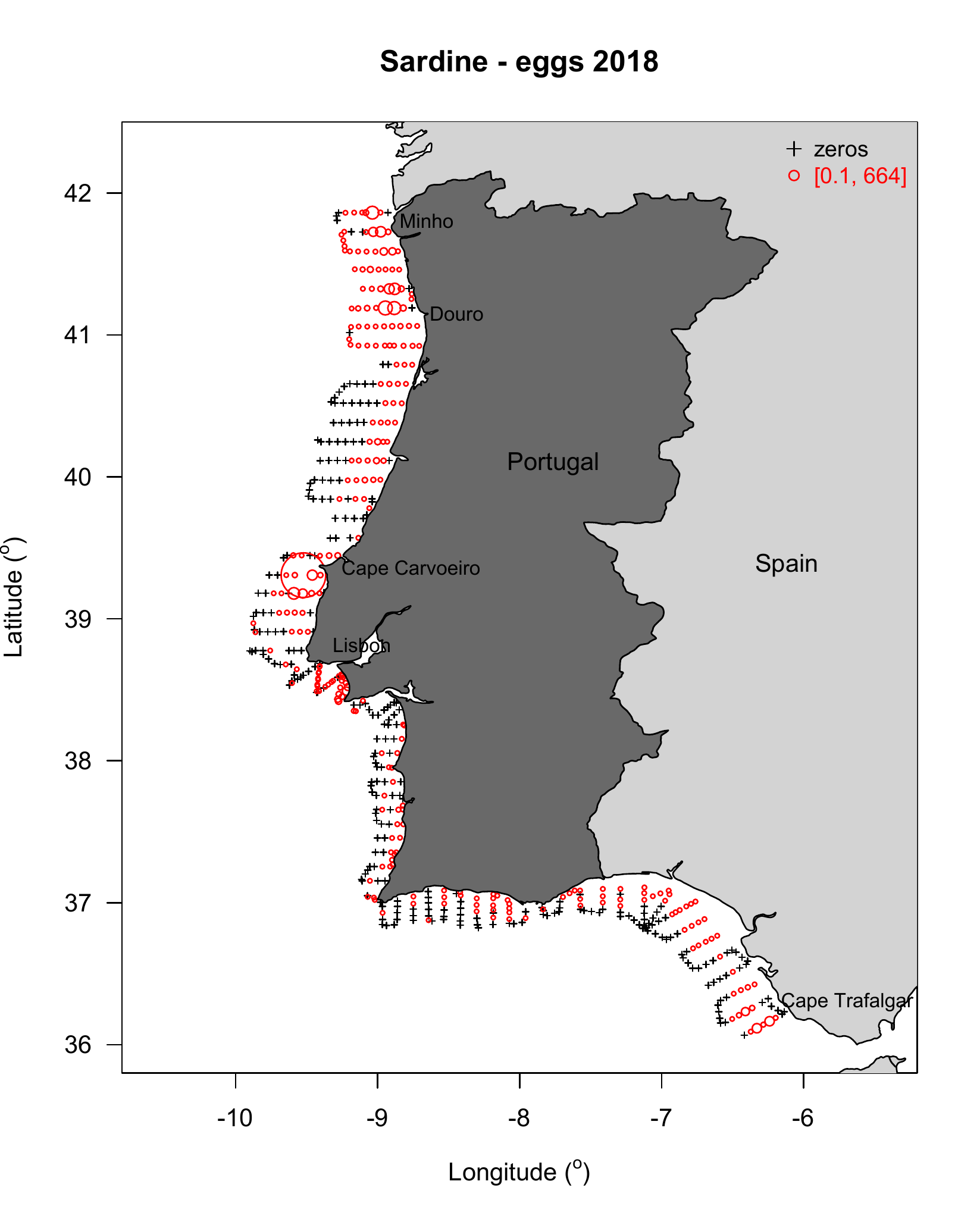}
	\caption{Observations of sardine eggs per $m^3$ collected in spring 2018 during IPMA PELAGO survey covering the Portuguese continental coast and the Spanish waters of the Gulf of Cadiz. Egg densities are proportional to circles area and egg absence denoted by crosses.}
	\label{data}
\end{figure}

The regular sampling design of these campaigns consists of diurnal echo-sounder recording (and fishing hauls) along transects, perpendicular to the shore line and spaced 8 nmi. Concurrently ichthyoplankton samples are collected, from water pumped continuously from 3 m depth, using a Continuous Underway Fish Egg Sampler (CUFES) \citep{Checkley1997}. The samples are obtained at every 3 nmi, corresponding to an integrated water volume filtered of around 11 $m^3$ per sample.

Together with the information on egg abundance (production) and spawning area definition, environmental variables such as temperature, salinity and fluorescence (proxy for chlorophylla concentration) are obtained continuously by sensors associated to the CUFES system while depth is recorded by the scientific echo-sounder. These covariates are available at the egg observation locations.

The samples are preserved onboard, with a formaldehyde solution at 4\% in water. After the survey in the laboratory, all ichthyoplankton organisms are sorted and the principal species (commercially exploited) are identified and counted. Egg densities are calculated using the volume of water filtered per sample.

During the 2018 PELAGO survey a total of 553 CUFES samples were collected and analysed. Sardine eggs were present in 49\% of the samples with densities varying from 0.1 to 664 eggs/$m^3$. The eggs were distributed almost over the whole area of the continental shelf surveyed with higher densities in the south, in the eastern region of the Gulf of Cadiz, in the central western coast, south of Cape Carvoeiro and in the more northern shelf, in the region between the rivers Douro and Minho (Figure 1.) 
The spatial resolution and area coverage of these regular observations is very high for open ocean biological sampling however, the biology/ecology of the species and the regional oceanography lead to a patchy distribution of the pelagic eggs and also a considerable inter-annual variability \citep{Bernal_2007, uriarte, masse}. Consequently, the data structure usually shows a high percentage of zeros (egg absence) and some extreme values (spawning hot spots).



\begin{figure}[ht]
	\centering
	\includegraphics[width=12cm]{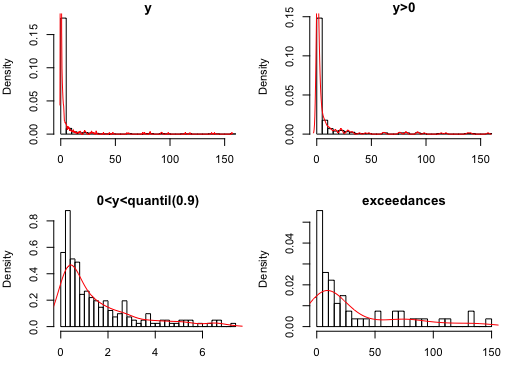}
	\caption{Histograms for: all egg data ($y$), positive data ($y>0$), data between 0 and 90\% quantile, exceedances above 90\% quantile}
	\label{hist}
\end{figure}

As we can see in Figure \ref{hist}, the data structure shows a high percentage of zeros, and extreme values.


We start by illustrating the zero-inflated model to deal with the high percentage of zeros in the data.  

\begin{figure}[h]
	\centering
	\includegraphics[width=5.5cm]{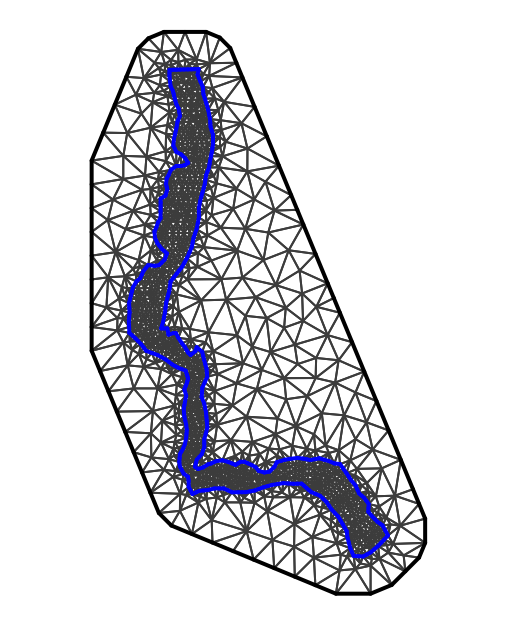}
	\caption{Mesh covering the study region off the continental Portuguese coast and the Spanish shelf in the Gulf of Cadiz.}
	\label{mesh}
\end{figure}

For the SPDE approximation, we use the mesh plotted in Figure \ref{mesh}.

In both equations (\ref{modBIN}) and (\ref{modGAMMA}), we consider as possible covariates $\{X_{i,m}(s)\}$  depth, temperature, salinity and fluorescence. Since we require covariate values in both observations locations and mesh nodes, a spatial extrapolation of the environmental covariates was previously made using a non-parametric model (additional details are provided in Appendix A).

Depth seems to be significant to explain the variation in the probability of occurrence, whereas fluorescence seems to be significant to explain both the probability of occurrence and the positive sardine eggs density (Table \ref{coef}).

\begin{table}[ht]
	\centering
	{\begin{tabular}{lrrrrr}
			\hline
			\cline{1-6}
			Coefficient &  mean  &   sd & Q0.025 & Q0.5 & Q0.975 \\
			\hline
			$\alpha_1$  &  -0.34 & 0.79  &  -2.04  &-0.31   &  1.18   \\
			$\alpha_2$  &  -0.06 &0.26   & -0.60  &-0.06    & 0.43    \\
			z.depth & 0.32 &0.12   &  0.08  & 0.32    & 0.55     \\
			y.depth & 0.09 &0.06  &  -0.03  & 0.08    & 0.20     \\
			z.temp  & 0.39 &0.53  &  -0.57  & 0.35    & 1.55   \\
			y.temp & -0.08 &0.22  &  -0.50  &-0.08   &  0.37    \\
			z.sal   & 0.20 &0.19  &  -0.16   &0.20    & 0.57   \\
			y.sal   & 0.08 &0.15   & -0.22   &0.08   &  0.39    \\
			z.fluor & 0.52 &0.22   &  0.09  & 0.52   &  0.97   \\
			y.fluor  & 0.23 &0.10    & 0.03   & 0.23   &  0.43    \\
			\hline										
	\end{tabular}}
	\caption{Regression coefficients for the zero-inflated model. Posterior mean, standard deviation and relevant posterior quantiles (Qp represents the quantile of probability p, so we represent the median and the 95\% credible intervals)}
	\label{coef}
\end{table}

The posterior mean is obtained at mesh nodes locations, from where a projection for a grid of 1km x 1km cells was made.


Our target of interest is the posterior mean of the sardine eggs density, $E[Y(s) \mid \theta_2]$. Following the law of total expectation, it follows that $E[Y(s) \mid \theta_2]=E[E[Y(s) \mid Z(s), \theta_2]]$. Since

$$E[Y(s) \mid Z(s), \theta_2]=\left\{\begin{array}{rc}
E[Y(s) \mid Z(s)=1, \theta_2],&\mbox{with probability}\quad p(s),\\
E[Y(s) \mid Z(s)=0, \theta_2], &\mbox{with probability}\quad 1-p(s).
\end{array}
\right.
$$
it results that $E[Y(s) \mid \theta_2]=E[Y(s) \mid Z(s)=1, \theta_2] p(s)$. Thus, here the estimates of the average number of sardine eggs by $m^3$ in a grid cell $c$ will be obtained by the product between the mean probability of occurrence and the average number of sardine eggs conditional on eggs being present in that cell.

\begin{figure}[ht]
	\centering
	\includegraphics[width=12cm]{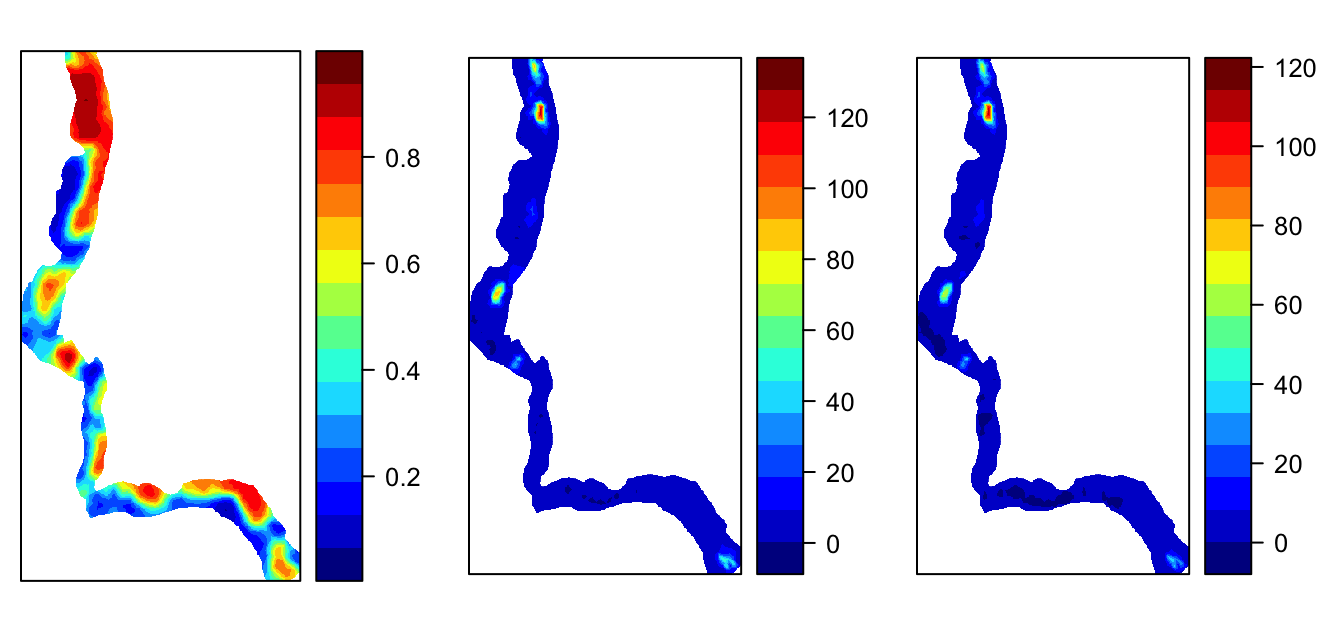}
	\caption{Posterior mean of: probability of presence (left), density conditional to $Z(s)=1$ (middle, eggs/$m^3$) and density (right, eggs/$m^3$)}
	\label{map1}
\end{figure}

\begin{figure}[ht]
	\centering
	\includegraphics[width=8cm]{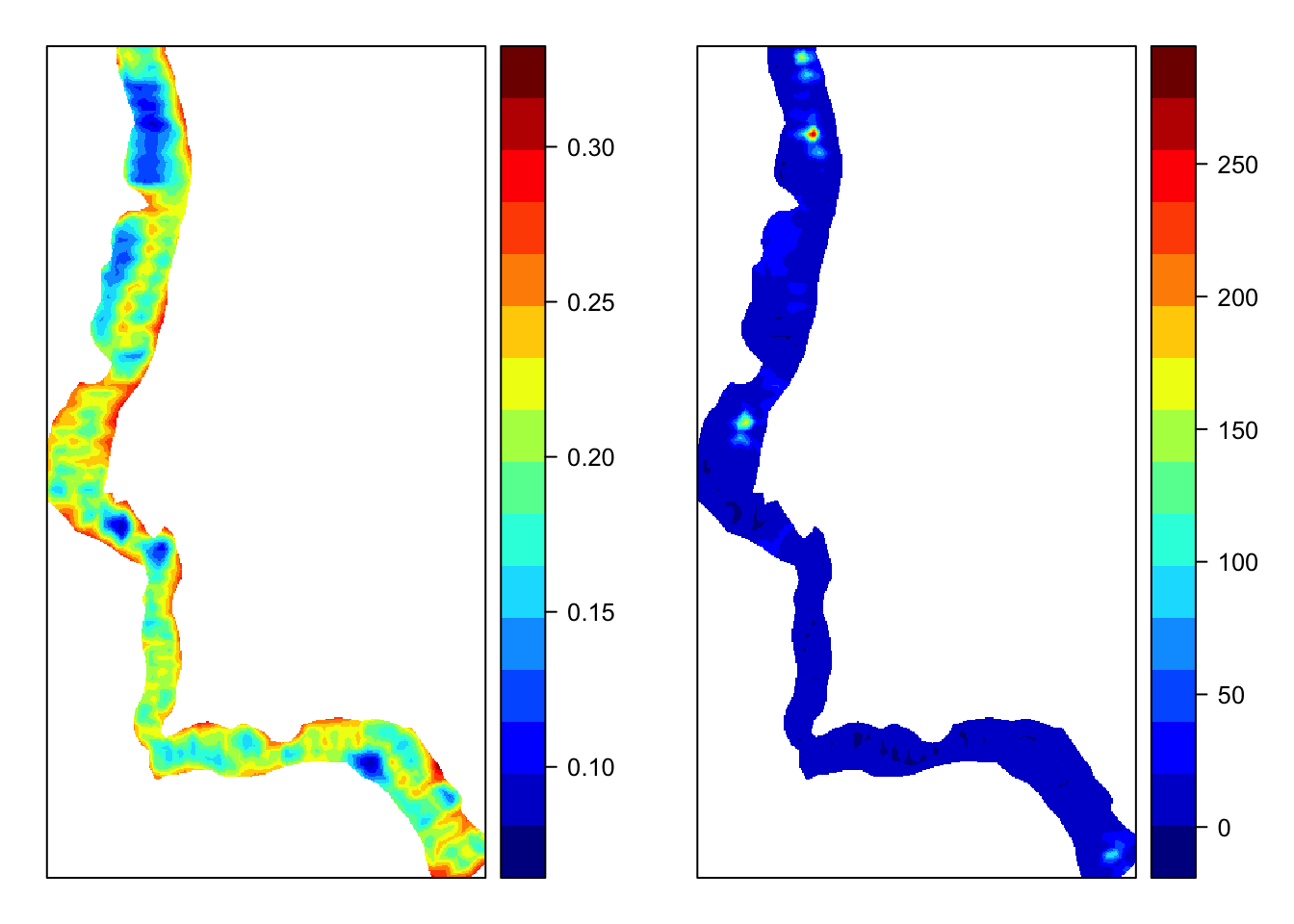}
	\caption{Standard deviation of: probability of presence (left), and density conditional to $Z(s)=1$ (right, eggs/$m^3$)}
	\label{map2}
\end{figure}

Figures \ref{map1} and \ref{map2} show the posterior mean of probability of occurrence, posterior mean of sardine eggs density conditional to the occurrence, and estimates of sardine eggs density. Note that the variability is very high in some cells (Figure \ref{map2}). The prediction obtained in these areas may be questionable. The high estimates are probably influenced by the extreme values in the sample.

\begin{figure}[ht]
	\centering
	\includegraphics[width=8cm]{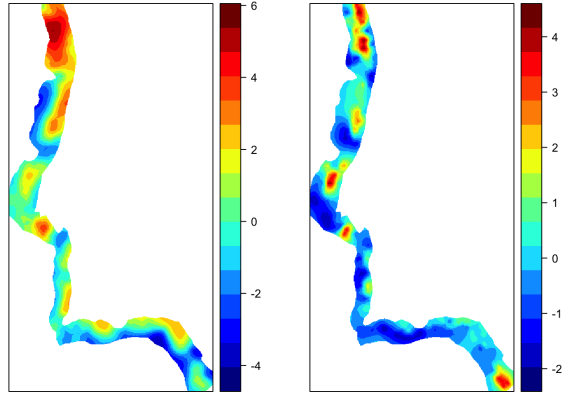}
	\caption{Posterior mean of: random field $W_1(s)$ (left) and $W_2(s)$ (right) defined in equations (\ref{modBIN}) and (\ref{modGAMMA}).}
	\label{map3}
\end{figure}

\begin{figure}[ht]
	\centering
	\includegraphics[width=8cm]{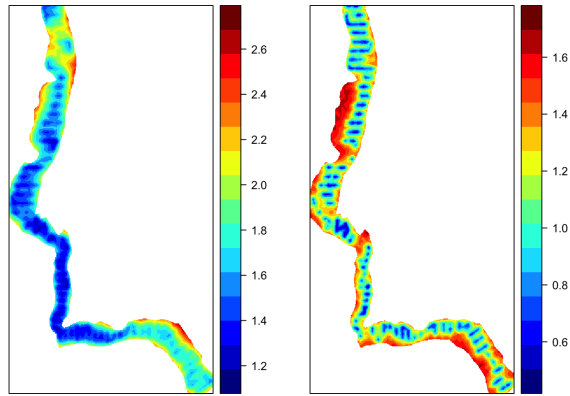}
	\caption{Standard deviation of: random field $W_1(s)$ (left) and $W_2(s)$ (right) defined in equations (\ref{modBIN}) and (\ref{modGAMMA}).}
	\label{map4}
\end{figure}

Figures \ref{map3} and \ref{map4} show the posterior mean of the random fields $W_1(s)$ and $W_2(s)$, defined in equations (\ref{modBIN}) and (\ref{modGAMMA}), and the respective standard deviation. $W_i(s)$ represents a spatially structured random effect, which considers local variability not taken into account by the covariates. Naturally, the standard deviation is lower along the transects defined in the sampling design and it is higher at the border of the study domain. 

To improve the estimates accuracy in cells with high variability, due to the extreme values, we adjust a geostatistical Generalised Pareto model to the exceedances over a threshold $u$.

\begin{figure}[ht]
	\centering
	\includegraphics[width=8cm]{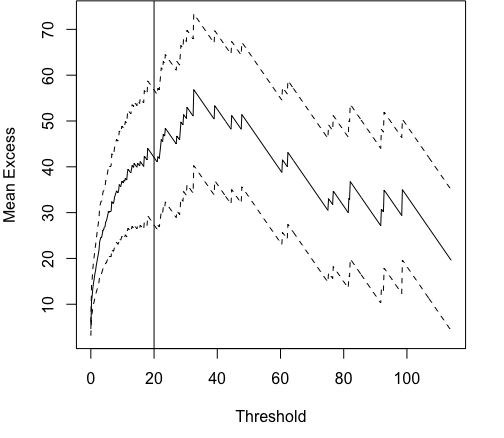}
	\caption{Mean residual life plot for threshold selection.}
	\label{th}
\end{figure}

\begin{figure}[ht]
	\centering
	\includegraphics[width=10cm]{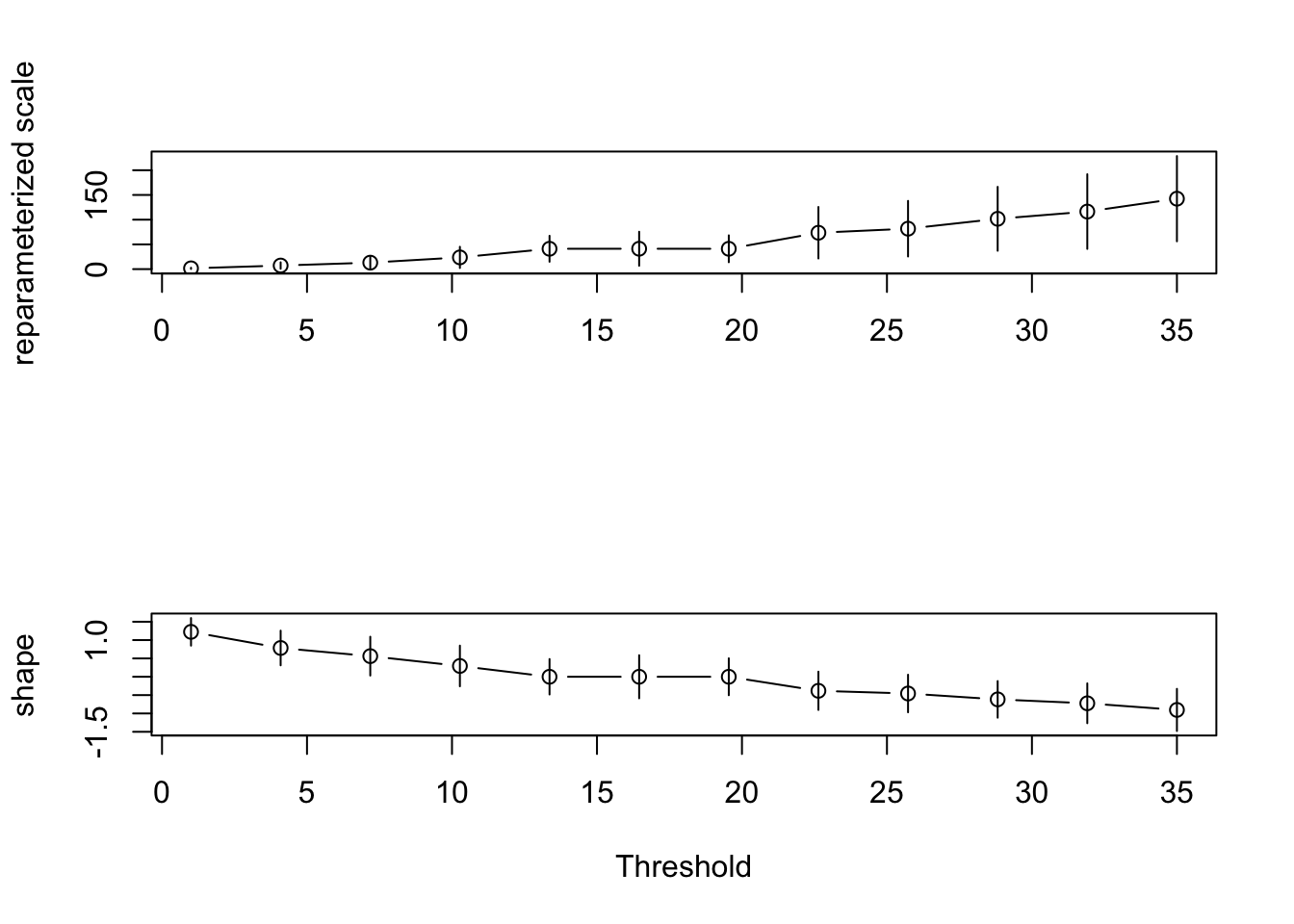}
	\caption{GP parameter estimates at a range of thresholds.}
	\label{th2}
\end{figure}

In this case, looking at the mean residual life plot (Figure \ref{th}), it seems that a good candidate for the selected threshold is around 35 since above that value the plot is approximately linear in $u$. However, only 17 observations are above that threshold, which can lead to high variability. Figure \ref{th2} presents the estimates for the shape and the standardized scale obtained by fitting a Generalised Pareto (GP) distribution for the exceedances at a range of thresholds between 1 and 35. Note that the variability increases significantly for thresholds above 20. Here, we choose a threshold of 20 (the original data have 30 observations above that value).

Among the possible covariates $\{X_{i,m}(s)\}$ being considered in equations (\ref{modBIN2}) and (\ref{modGP}), none seems to be significant to explain the probability of exceeding the chosen threshold (Table \ref{coef_extremes}).

\begin{table}[ht]
	\centering
	{\begin{tabular}{lrrrrr}
			\hline
			\cline{1-6}
			Coefficient &  mean  &   sd & Q0.025 & Q0.5 & Q0.975 \\
			\hline
			$\alpha^*_1$  &   -7.36 &3.40 &  -14.68 & -7.18  &  -0.95 \\
			$\alpha^*_2$   & 3.00 &0.38  &   2.26 &  3.00  &   3.75 \\
			$z^*.depth$ & 0.28 &0.61 &   -0.93 &  0.28   &  1.48  \\
			$y^*.depth$ & 0.10 &0.18  &  -0.26 &  0.10  &   0.46  \\
			$z^*.temp$  & -0.59 &3.21 &   -6.87 & -0.66  &   6.07  \\
			$y^*.temp$ & 0.14 &0.42 &   -0.68  & 0.14  &     0.96 \\
			$z^*.sal$   & 1.37 &2.06  &  -2.61 &  1.33  &   5.54  \\
			$y^*.sal$  &  0.07 &0.41  &  -0.73 &  0.07  &   0.88   \\
			$z^*.fluor$ & 0.35 &1.17  &  -1.94  & 0.34  &   2.69  \\
			$y^*.fluor$  & 0.33 &0.24  &  -0.15 &  0.33  &   0.81  \\
			\hline										
	\end{tabular}}
	\caption{Regression coefficients for the extremes model. Posterior mean, standard deviation and relevant posterior quantiles (Qp represents the quantile of probability p, so we represent the median and the 95\% credible intervals)}
	\label{coef_extremes}
\end{table}

\begin{figure}[ht]
	\centering
	\includegraphics[width=12cm]{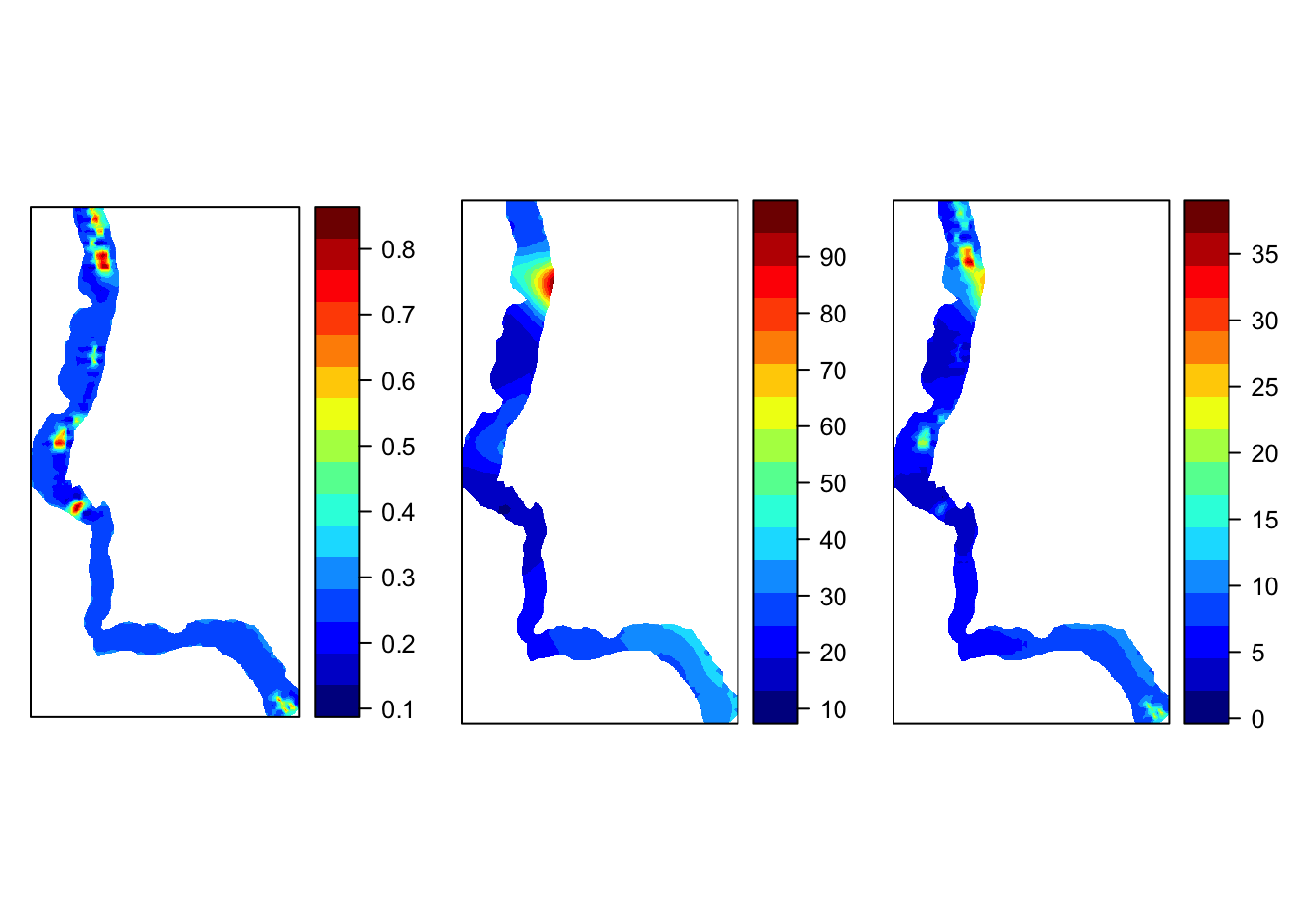}
	\caption{Posterior mean of: probability of exceeding the threshold (left), exceedances conditional to $z^*(s)=1$ (middle, eggs/$m^3$) and exceedances (right, eggs/$m^3$)}
	\label{map5}
\end{figure}

\begin{figure}[ht]
	\centering
	\includegraphics[width=8cm]{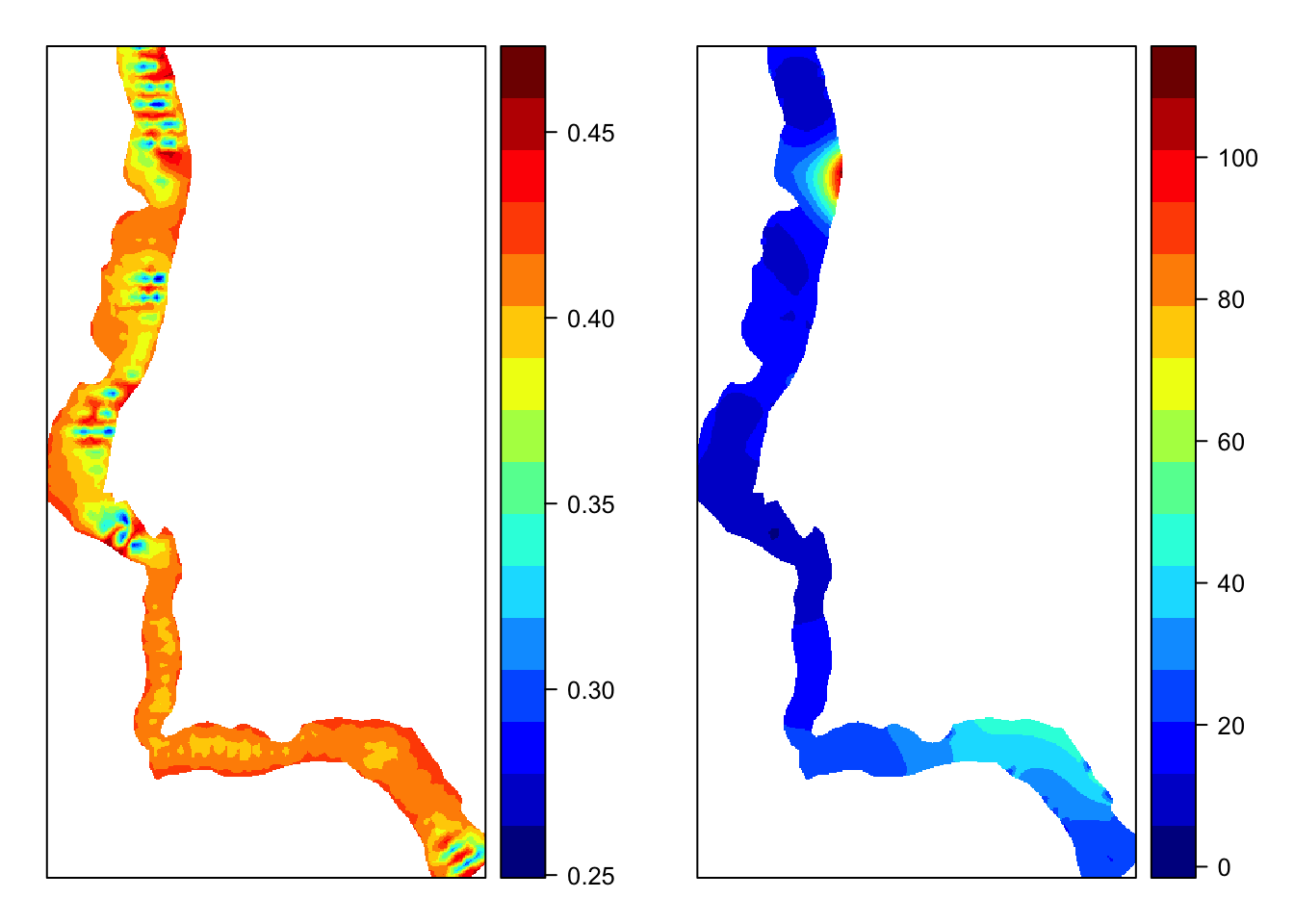}
	\caption{Standard deviation of: probability of exceeding the threshold (left), and exceedances conditional on $Z^*(s)=1$ (right)}
	\label{map6}
\end{figure}

The posterior mean of exceedance probability, the conditional posterior mean of exceedances and the estimated exceedances are represented in Figure \ref{map5}. Moreover, the respective standard deviations are represented in Figure \ref{map6}. Note that the standard deviation is now much lower than those obtained using Model I. 

\clearpage

\subsection{Combining the models for zero-inflated and extreme values}
\label{jointmodel}

To combine the 
advantages of both Model I (zero-inflated) and Model II (for extremes) to estimate the sardine eggs density there are different possible ways to proceed. Here three different approaches are proposed:

\begin{enumerate}[label=(\alph*)]
	
\item[(A)] The estimates are defined according to Model I, except where they are higher than the chosen threshold u. In those cells, the estimates are replaced by the values derived from Model II (estimated exceedances plus threshold). The mathematical representation becomes


$$\hat{D}(s)=\left\{\begin{array}{rc}
E\left[Y(s) \mid \theta_2\right],&\mbox{if}\quad E\left[Y(s) \mid \theta_2\right] \leq u,\\
E[Y^*(s) \mid \theta^*_2]+u, &\mbox{if}\quad E[Y(s) \mid \theta_2] > u.
\end{array}
\right.
$$

\item[(B)] The estimates are defined according to Model I, except where the probability of exceeding the chosen threshold u is higher than 0.5. In those cells, estimates are replaced by the values derived from Model II. The mathematical representation becomes


$$\hat{D}(s)=\left\{\begin{array}{rc}
E[Y(s) \mid \theta_2],&\mbox{if}\quad E[p^*(s) \mid \theta^*_1] \leq 0.5,\\
E[Y^*(s) \mid \theta^*_2]+u, &\mbox{if}\quad E[p^*(s) \mid \theta^*_1] > 0.5.
\end{array}
\right.
$$
	
\item[(C)] The estimates are defined according to a weighted average of Model I's estimates and Model II's estimates, where the weights correspond to the probability of exceeding the threshold u and its complementary probability. The mathematical representation becomes


$$\hat{D}(s)=E[p^*(s) \mid \theta^*_1](E[Y^*(s) \mid \theta^*_2]+u)+(1-E[p^*(s) \mid \theta^*_1])E[Y(s) \mid \theta_2] $$

\end{enumerate}

\begin{figure}[h]
	\centering
	\includegraphics[width=12cm]{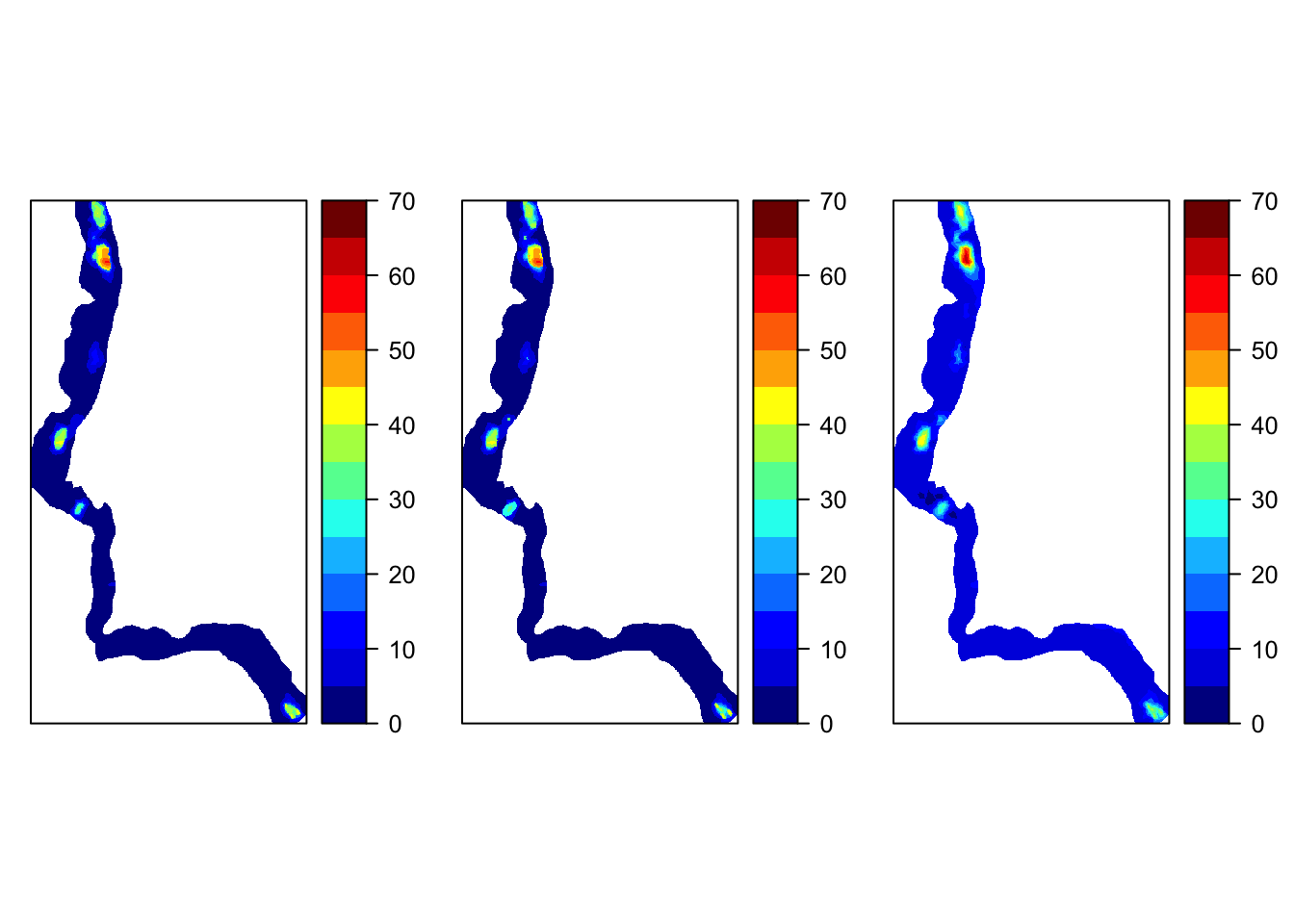}
	\caption{Estimated density of sardine eggs (eggs/$m^3$): using approach (A) (left), approach (B) (middle) and approach (C) (right) given in section \ref{jointmodel}}
	\label{map7}
\end{figure}

A comparison between the estimation results derived from approaches (A), (B) and (C) is represented in Figure \ref{map7}. Due to the weighted estimation approach (C) offers a smoother spatial variation for the density estimation of the sardine eggs from the 2018 survey in the Portuguese coast and Spanish waters of the Gulf of Cadiz.

The choice of the most adequate approach may depend on the application itself. In generic terms, we believe that for most cases the suggested approach (C) will bring more information into the model, as it takes into account the estimated probabilities of exceeding a pre-defined threshold, not being dictated by a binary decision.


\clearpage


\section{Discussion}\label{disc}


Since the early 1990s, the importance of geostatistics has been widely recognised in fisheries and marine ecology. It has proven to be a useful tool for sampling designing and estimation of biomass and abundances and their precision, for populations around the world \citep{ICES1993, Rivoirard2000}. Unfortunately, due to the inherent difficulties of modelling fisheries data, there are still many open problems in the construction of accurate distribution maps, which are hard to be tackled through classical geostatistical tools. These difficulties include the need to consider 
unobserved sources of variability, 
or space and time dependent characteristics, which are easier to be handled by adopting Bayesian mixed modelling approaches. 
In particular, the problems discussed in this work, excess of zeros or presence of extremes values, so common in the study of population dynamics, could be mitigated by adopting non-linear techniques, like indicator or disjunctive kriging \citep{Rivoirard1994}. The topcut model proposed in \cite{Rivoirard2013} could also provide a valuable substitute to linear kriging in the case of a skewed distribution with a few high values. Yet, these classic kriging approaches find it difficult to compete with the extra flexibility offered by Bayesian INLA approaches, which allow for integrated solutions for all previously described problems, at an acceptable computational cost.

The hierarchical model here proposed essentially combines separate submodels for each of the data components, the zeros, the extremes, and the rest. While this might seem a natural approach, allowing inferences about separate ecological processes, potentially governed by different covariates, we are unaware of the approach having been used in the ecological literature. See for instance Martínez-Minaya et al (2018) on ``state-of-the-art'', where zero-inflated approaches are discussed but nothing is said about extremes nor about a combination of these issues. 

The approach here presented is a first development dealing with these type of biological data and was at this stage applied for a single survey. Naturally, it seems possible to extend this spatial model to incorporate sampling across years, ending up with a spatio-temporal model. Given the high inter-annual variability of sardine egg density distribution, for a more comprehensive study of its dynamics further surveys need to be added to the analyses and the temporal dimension should be considered. Such  implementation is being developed.

We presented a sensitivity analysis to the choice of threshold in the extreme value model, and based on it chose a value of 20 for our illustrative example. This choice was however somewhat arbitrary, and this is an area that deserves further investigation, since the results might be sensitive to the threshold choice. This is a new problem compared to the zero-inflation component, where the threshold as a natural definition (presence versus absence). In the case of extremes defining the threshold beyond which observations are modelled as extremes becomes a model selection question.

There are a few natural and conceptually straightforward extensions to our methodological proposal. One might introduce a non-linear relation between environmental covariates and the variable of interest (generalised additive mixed models), and exploring different families for the distributions considered for the variable of interest, among the exponential family. Finally, exploring how different covariates might explain each of the different components of the model seems like a possible way to make inferences about different ecological sub-processes affecting the overall distribution of sardines. 

Since our method depends on the choice of a threshold, we intend to extend this approach in future work, using an extended Generalised Pareto distribution as proposed in \cite{naveau2016}. In addition, to avoid a threshold selection, this model allows to model simultaneously both the bulk and the tail of the distribution.


Ecological data is typically messy, and standard models are often not adequate to deal with them, presenting a lack of flexibility that prevents fully efficient inferences. Here we present an approach that might be used to deal with situations where not only there are too many zeros in the data, but also there might be a proportion of very large values. Coping adequately with the large values is fundamental, and might be of practical consequence. If as in the example presented, one is interested in predicting say a total that is effectively a sum over space, a small number of large observations might correspond to a non negligible proportion of the entire population, and hence, not only we have a bad model goodness-of-fit, but we also might observe considerable bias. On the other hand, localised high abundance values, and zeros, as the present case study shows, have biological/ecological meaning (eg. spawning hot spots and less favourable areas for spawning or species absence) and can not be overlooked. Hence, the models should be able to incorporate all observations and at the same time perform adequately to provide good estimates for the components of the real ecosystems.
While we illustrated our models using sardine eggs, there is nothing specific regarding the approach that would prevent its utilization for other scenarios. One could use it to model say terrestrial insects, or plants, or even non ecological data. We have illustrated  that besides having to handle carefully the high proportion of zeros in some data sets, an issue which has been abundantly discussed in the literature \citep[e.g.][]{Martin2005},  it is also important to consider the extreme values. These could represent different processes affecting a subset of the data. As we do so we move further away from a traditional view of modelling where extreme observations would be called outliers to a world where we appreciate that extreme observations might contain useful information worth modelling.

\section*{Acknowledgments} \label{acknowledgements}

This work is partially financed by national funds through FCT – Fundação para a Ciência e a Tecnologia under the projects UIDB/00006/2020, UIDB/04050/2020, PTDC/MAT-STA/28243/2017 and PTDC/MAT-STA/28649/2017. The survey data analysed was collected under the framework programme PNAB: Portuguese Marine Surveying Programme - P03M02 (EU Data Collection Framework EU-DCF, FEAMP), and the current work was developed within the scope of project SARDINHA2020 - Ecosystem approach towards a sustainable sardine fishery exploitation (Mar2020-MAR-01.04.02-FEAMP-0009).



\bibliographystyle{asa2} 
\bibliography{eggs}

\appendix

\section{Spatial kernel smoothing on environmental covariates}\label{kern}




Although we do not necessarily expect a smooth spatial variation in the sardine eggs density, we expect that behaviour for the environmental covariates. Moreover, the proposed spatial model for the response of interest requires the covariates evaluation at specific mesh nodes locations. Thus, as a precursor to the analysis itself, we have implemented an extrapolation of the covariates from the observed locations to the entire study domain, using a non parametric method.

The idea behind this method, also known as the Nadaraya-Watson smoother \citep{M.P.Wand1994} is the following: if the observed values are $y(s_1),...,y(s_n)$ at locations $s_1,...,s_n$ respectively, then the smoothed value at a location $u$ can be given by

\begin{equation}
g(u)=\frac{\sum_i k(u-s_i) y(s_i)}{\sum_i k(u-s_i)}
\end{equation}
\noindent where $k$ is a probability density. A common choice for the density is the Gaussian kernel.

\begin{figure}[ht]
	\centering
	\includegraphics[width=13cm]{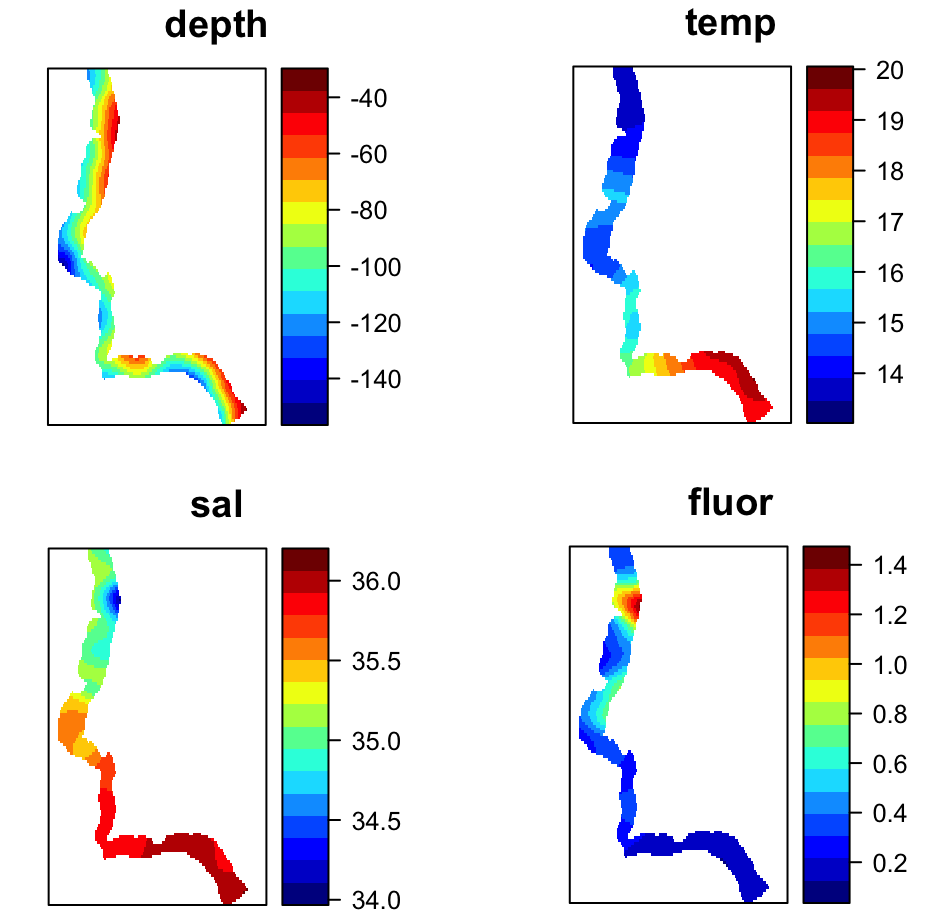}
	\caption{Extrapolated covariates by a spatial kernel smoothing method}
	\label{cov}
\end{figure}

The spatial smoothing on covariates shows that in general, the temperature and salinity are higher in the south of the study area, whereas depth is, as naturally expected, lower near the coastline and the fluorescence was higher in the NW shelf associated to river outflow and/or upwelled waters (Figure \ref{cov}).

\clearpage

\end{document}